\documentclass[11pt,fleqn]{article}
\usepackage{psfig}
\begin{document}

\title{Possible Superfluidity \\ in thin corrugated Annulus}
\author{Zotin K.-H. Chu} 
\date{  
WIPM, 30, Xiao-Hong Shan West, Wuhan 430071, PR China and
\\ Math.-Phys. Centre, 4-601, Building C,  Beijingcheng,
Baixingkangcheng, \\ Road Changsha, Wulumuqi 830013, China}
%
\maketitle
\begin{abstract}
We revisit the persistent flow of a superfluid in a thin
wavy-rough annulus. The existence of a phase memory around this
thin corrugated annulus is shown to be responsible for the energy
minima with a periodic dependence on the total momentum which is
directly related to the quantization of circulation. We also
illustrate the general features using the ideal Bose gas  as an
example.
\end{abstract}
\doublerulesep=6.5mm        
\baselineskip=6.5mm \oddsidemargin-1mm
%
\section{Introduction} One of the most remarkable properties of
macroscopic quantum systems is the phenomenon of persistent flow.
In a superconductor, persistent flow is electrical current without
resistance: current in a loop of superconducting wire will flow
essentially forever [1]. In a superfluid such as liquid helium
below the lambda point, the frictionless flow allows persistent
circulation in a hollow toroid [2-3]. Although the dynamics of
superfluids follows quite directly from the simple assumption that
the quantum field of the particles has a mean value which may be
treated as a macroscopic variable [4] (Anderson mentioned that it
is as legitimate to treat the quantum field amplitude as a
macroscopic dynamical variable as it is the position of a solid
body; both represent a broken symmetry which, however, cannot be
conveniently repaired until one gets to the stage of quantizing
and studying the quantum fluctuations of the macroscopic behavior
of the system).\newline To help researchers to easily understand
the superfluidity in the ring-type geometry, in this short paper,
we only consider the uniform flow around a circular wavy-rough
(thin) annulus with the linear dimension of the cross section
smaller than the mean radius ($R=(r_i+r_o)/2$). The amplitude of
the wavy-rough corrugations ($\epsilon$) is presumed to be much
smaller than  the linear dimension of the cross section (say,
$r_o-r_i$ and $r_i$ or $r_o$, cf. Fig. 1; i.e. $\epsilon \ll
r_o-r_i \ll R$).
\section{Theoretical Formulations}
Let $x$ be a coordinate along the average ring of the wavy-rough
thin annulus ($x/R$ representing the angle variable). The explicit
dependence on the two additional coordinates required to locate an
atom within the cross section is neglected here for simplicity.
The system of $N$ atoms can be described by a symmetric wave
function $\psi (x_s)$ of the $N$ variables $x_s$ together with the
conjugate momenta $p_s$ $(s=1,\cdots,N)$. The interaction with the
walls shall be neglected here so that the total momentum $P=
\sum_s p_s$ (or the total angular momentum $PR$) is a constant of
the motion.  \newline We can obtain a solution
\begin{equation}
 \psi (x_s)= \chi (x_s - x_{s'}) \exp[{{i P(\sum_s x_s)}/{N
 \hbar}}],
\end{equation}
of the Schr\"{o}dinger equation
\begin{displaymath}
{\cal H} \psi=E \psi.
\end{displaymath}
Here,
\begin{displaymath}
P \chi=0
\end{displaymath}
 and $s,s'=1,\cdots,N$.
 In
fact, we have
\begin{equation}
 p_s = \frac{\hbar}{i} \frac{\partial}{\partial x_s}
\end{equation}
which only come from the kinetic energy $\sum_s p_s^2/2m$. From
above equations and/or mathematical expressions, the substitution
for $\psi$ into the  Schr\"{o}dinger equation gives
\begin{equation}
 {\cal H} \chi =  e \chi
\end{equation}
together with the total energy
\begin{equation}
 E = \frac{P^2}{2 M}+e,
\end{equation}
where $M=N m$ is the total mass of the system. Note that it is
customary to separate the motion of the center of gravity ($\sum_s
x_s)/N$. Thus, $e$ which is due to the relative motion of the
atoms, is normally independent of the total momentum $P$. \newline
Nevertheless $\psi$ should be single valued for superfluid systems
and must return to the same value when an atom is brought around
the ring to its original position. This means whenever $x_s = x_s
+2\pi R$, we have
\begin{displaymath}
\chi =\chi \,f
\end{displaymath}
 with
\begin{displaymath}
 f=\exp[-2\pi i(PR/N
\hbar)].
\end{displaymath}
Hence, $e$ depends upon $P$ in the superfluid state. However, as
$\chi$ depends only on the differences of $x_s$ and then $\chi$
remains the same if all $N$ of them are increased by $2\pi R$.
This also requires that $f^N=1$ and $P=k \hbar/R$ where $k$ is an
integer. The latter confirms that the eigenvalue of the total
angular momentum $PR$ to be integer multiples of $\hbar$.
\newline With above reasoning, we also have, for the eigenvalue
$e$, $e(P+N \hbar/R)=e(P)$ or $e(P)$ is periodic with $N\hbar/R$.
Moreover, as a reverse rotation cannot affect the energy $E$, the
same holds for $e$ so that $e$ is an even function of $P$. To be
explicit, a stationary state of the system can be characterized by
$P$ and an additional set of $N-1$ quantum numbers, say, $n$. The
corresponding eigenvalue of the energy is then of the general form
\begin{equation}
 E_n (P)=\frac{P^2}{2M}+e_n (P),
\end{equation}
with
\begin{equation}
 e_n (P+\frac{N\hbar}{R})=e_n (P), \hspace*{12mm} e_n(-P)=e_n (P).
\end{equation}
A dependence of the energy on the total momentum through $e$
indicates a phase memory which, irrespective of its origin,
extends around the whole thin wavy-rough annulus.
\section{Persistent Flow at T=0} We only consider the conditions
at the absolute zero of temperature (T=0) for simplicity. It means
$P=0$ and the ground state of the system will be characterized
also by (the set of additional quantum numbers) $n=0$. Now,
\begin{equation}
 E_0 (P)=\frac{P^2}{2M}+e_0 (P)
\end{equation}
represents the lowest value of the energy for a given $P$ and
$e_0(P)$ has a minimum at $P=0$ (and this minimum is periodically
repeated : $e_0 (P+N\hbar/R)=e_0(P)$). If we presume $e_0(P)$ has
a finite slope as $P\rightarrow 0$ from either side, then it can
be illustrated schematically in Fig. 2. Similarly, as shown in
Fig. 3, $E_0(P)$ also exhibits minima when $P$ is an integer
multiple of $N\hbar/R$ with an absolute magnitude below a {\it
critical} value $P_c$. Next, we need to consider the interaction
of the system with the (container) walls along (with) the ensuing
equilibrium to understand the significance of these minima. After
starting from an arbitrary initial state, equilibrium can be
reached by transitions involving an exchange of momentum and
energy. Once the walls (of the container) are held at zero
temperature, the system will be brought to the lowest state and
hence to a vanishing momentum (if, as in normal cases, there are
no other minima of $E_0$). \newline If, however, there are  other
minima the system can reach one of them through a rapid succession
of transitions, each involving an energy loss accompanied by a
small transfer of momentum to the walls. From this on, only those
transitions could occur in which the energy further decreases with
a momentum transfer comparable to or larger than the
macroscopically large value $N\hbar/R$. Such transitions can thus
be safely considered to be so highly rare as to cause a {\it
metastability} which, in effect, will let the system remain in a
state with (momentum) $P_{\mu}=\mu N \hbar/R$ ($\mu$ is an
integer). \newline In view of the general definition of the drift
velocity $u=P/M$ ($M=N m$), the system therefore exhibits {\it
persistent flow} with (the drift velocity)
\begin{equation}
u_{\mu}=\frac{\mu \hbar}{m R}.
\end{equation}
Meanwhile the {\it circulation} defined by the line integral
around the ring is $2\pi R u_{\mu}=\mu (h/m)$. With this, above
just confirms the quantization with the {\it quantum of
circulation} $h/m$ (valid for $|P_{\mu}| \le P_c$).
\subsection{Critical Velocity}
From equation (7) we are concerned, in fact, no longer with a
minimum of $E_0$ at such large values of $P_{\mu}$ that $P^2/2M$
rises more steeply than $e_0$, because it allows a decrease of the
energy towards the low-momentum side. Using the derivatives
expression and $e'_0 (P_{\mu})=e'_0(0)$, we have a persistent flow
at the drift velocity $u_{\mu}$ requiring
\begin{equation}
 \frac{|P_{\mu}|}{M}=|u_{\mu}| < |e'_0 (0)|,
\end{equation}
so that the {\it critical} momentum is
\begin{equation}
 P_c= M  |e'_0(0)|  \hspace*{12mm} \mbox{or that} \hspace*{12mm}
 u_c=|e'_0(0)|
\end{equation}
represents a critical velocity. \newline The above pedagogical
statements or detailed explanations demonstrate the possibility to
derive some of the salient features of superfluidity from
fundamental principles. We shall give an example, an ideal Bose
gas, below to illustrate above mentioned point so that students
and researchers can gain more insights into its behavior.
\subsection{Example: Ideal Bose Gas}
We shall consider an ideal Bose gas for a pedagogical
illustration. In the interval $0\le P\le N \hbar/R$, the lowest
energy for a given value $P=\nu (\hbar/R)$ can be obtained by
assigning the momentum $p=\hbar/R$ to $\nu$ atoms and $p=0$ to the
($N-\nu$) atoms. We then have
\begin{equation}
 E_0(P)=\frac{\nu \hbar^2}{2 m R^2}=\frac{P \hbar}{2 m  R},
\end{equation}
and from equation (7) (as $M=N m$),
\begin{equation}
 e_0(P)=\frac{P\hbar}{2 m R}(1-\frac{P R}{N \hbar}).
\end{equation}
With other intervals  being almost the same (except the period
shift), we can determine the corresponding function and
demonstrate it by the dashed parabolic curves in Fig. 2  (setting
$N \hbar/R\equiv1$).
\newline Once we add $P^2/2M$ to above expression, we find (for
$P_{\mu} \le P \le P_{\mu+1}$)
\begin{equation}
 E_0(P)=\frac{P_{\mu}^2}{2M}+\frac{(P-P_{\mu})\hbar}{2 mR}
 (2\mu+1).
\end{equation}
The straight segments which connect the points on the parabola
$P^2/2M$ for $P=P_{\mu}$ and  $P=P_{\mu+1}$ represent this
function by the dashed line in Fig. 3 (setting $N
\hbar/R\equiv1$). As it has no other minimum than that at $P=0$,
we can conclude that the ideal Bose gas does not exhibit
persistent flow in a contained at rest. However, we remind the
readers that the above property derives from a uniform rotation,
rather than a translation, of the container (thin corrugated
annulus) so that the invariance of the relative velocity against
uniform uniform motion of the system of reference cannot be
invoked in concluding on the case of a contained at rest
considered here. \section{Conclusion} To conclude in brief, we
already  demonstrate the possibility to derive some of the
characteristic features of superfluidity from fundamental
principles in a weakly-corrugated thin annulus and we believe our
presentation will be useful to  researchers in relevant fields.


\newpage
\psfig{file=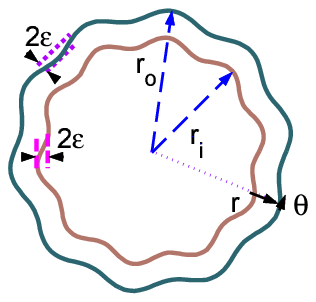,bbllx=1cm,bblly=18.5cm,bburx=14cm,bbury=25cm,rheight=6cm,rwidth=6cm,clip=}

\vspace{2mm}
\begin{figure}[h]
\hspace*{6mm} Fig. 1 \hspace*{1mm} Schematic plot of a thin
wavy-rough annulus. The amplitude \newline \hspace*{8mm} of the
corrugation $\epsilon \ll R=(r_i+r_o)/2$.
\end{figure}

\newpage
\psfig{file=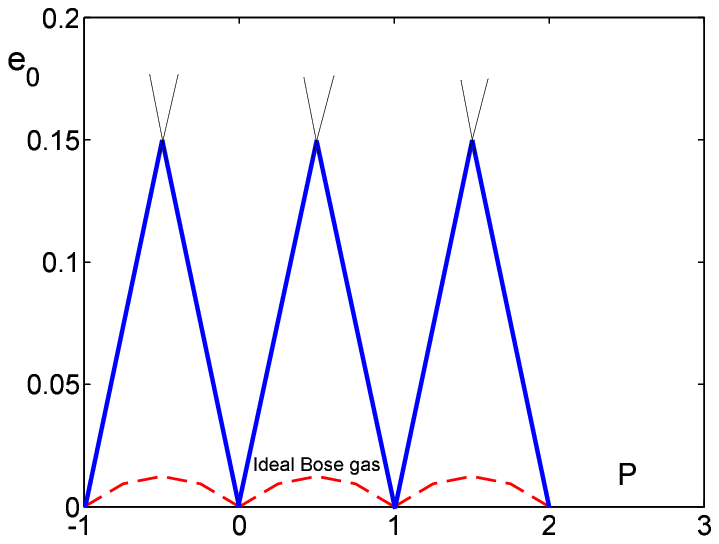,bbllx=0cm,bblly=19cm,bburx=10cm,bbury=26cm,rheight=8cm,rwidth=10cm,clip=}
\vspace{2mm}
\begin{figure}[h]
\hspace*{6mm} Fig. 2 \hspace*{1mm} Schematic illustration of $e_0
(P)$. It is presumed that finite slopes \newline \hspace*{8mm}
exist at the periodically repeating minimum. The dashed lines
\newline \hspace*{8mm} indicate the case of an ideal Bose gas.
$P$ is the total momentum. \newline \hspace*{8mm} Here, we set $N
\hbar/R\equiv1$.
\end{figure}
\newpage
\psfig{file=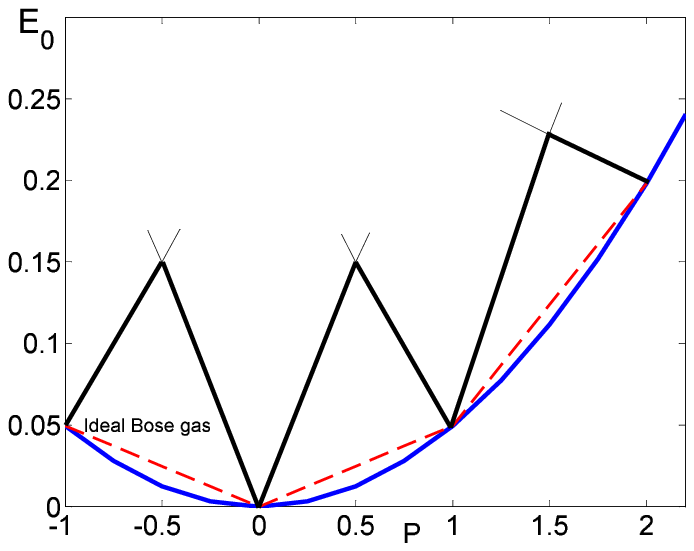,bbllx=0cm,bblly=19cm,bburx=10cm,bbury=26cm,rheight=8cm,rwidth=10cm,clip=}
\vspace{2mm}
\begin{figure}[h]
\hspace*{6mm} Fig. 3 \hspace*{1mm} Schematic illustration of $E_0
(P)$. The slope towards lower values \newline \hspace*{8mm} of
$|P|$ (for heavy solid lines) at successive minima decreases with
\newline \hspace*{8mm} increasing momentum (preventing the occurrence of further minima
\newline \hspace*{8mm} above a critical value $P_c$ of $|P|$). The dashed lines
indicate the case \newline \hspace*{8mm}  of an ideal Bose gas.
$P$ is the total momentum. Here, we set $N \hbar/R\equiv1$.
\end{figure}

\end{document}